\documentclass{article}
\usepackage{spconf,amsmath,graphicx}
\usepackage{algorithm}
\usepackage[table,xcdraw]{xcolor}
\usepackage[hang,flushmargin]{footmisc} 
\usepackage{booktabs}
\usepackage{amsfonts}
\usepackage{amssymb} 
\usepackage{bm,bbm}
\usepackage{hyperref}
\usepackage{multirow}
\usepackage{makecell}
\usepackage[noend]{algpseudocode}
\usepackage{pifont}
\usepackage{subcaption}

\usepackage{hyperref}

\title{DPM-TSE: A Diffusion Probabilistic Model for Target Sound Extraction}
\name{Jiarui Hai$^{1,\dagger}$\thanks{$^{\dagger}$Indicates equal contribution.}, 
Helin Wang$^{2,\dagger}$, 
Dongchao Yang$^{3}$, 
Karan Thakkar$^{1}$,
Najim Dehak$^{2}$,
Mounya Elhilali$^{1,2}$\thanks{This work was supported by ONR N00014-23-1-2050 and N00014-23-1-2086 and NIH U01AG058532}}
\address{
$^{1}$Laboratory for Computational Auditory Perception, Johns Hopkins University, Baltimore, USA\\
$^{2}$Center for Language and Speech Processing, Johns Hopkins University, Baltimore, USA\\
$^{3}$The Chinese University of Hongkong, Hongkong, China\\
}
\begin{document}
%
\maketitle
\begin{abstract}
Common target sound extraction (TSE) approaches primarily relied on discriminative approaches in order to separate the target sound while minimizing interference from the unwanted sources, with varying success in separating the target from the background. This study introduces DPM-TSE, a first generative method based on diffusion probabilistic modeling (DPM) for target sound extraction, to achieve both cleaner target renderings as well as improved separability from unwanted sounds. The technique also tackles common background noise issues with DPM by introducing a correction method for noise schedules and sample steps. This approach is evaluated using both objective and subjective quality metrics on the FSD Kaggle 2018 dataset. The results show that DPM-TSE has a significant improvement in perceived quality in terms of target extraction and purity.

\end{abstract}
\section{Introduction}
\label{sec:intro}
There are countless sounds in the world that offer crucial information about our environment, including the melody of a violin during a concert and sirens in the streets. Our daily lives could be significantly enhanced if we were able to create listening devices that could filter out unwanted sounds and focus on the sounds we want to hear. In recent years, machine hearing has studied target sound extraction and removal applications, which aim to identify specific speakers \cite{wang2019voicefilter}, musical instruments \cite{slizovskaia2021conditioned}, and sound events \cite{borsdorf2021universal,delcroix2021few,gfeller2021one,pishdadian2020learning}. Among them, the extraction of sound events is much more challenging than others because of a wide range of sounds, such as animal noises, baby cries, and telephone calls.
This work addresses the problem of target sound extraction (TSE).

TSE aims to separate the sound of a specific sound event class from a mixed audio given a target sound \cite{gfeller2021one,kong2020source,ochiai2020listen}. 
Researchers have explored the challenges of new classes and weakly-labelled data, with some proposing solutions such as combining one-hot-based and enrollment-based target sound extraction \cite{delcroix2021few}, weakly-supervised sound separation \cite{pishdadian2020learning}, and random sound mixing \cite{gfeller2021one}. 
These methods are based on discriminative models, 
which minimize the difference between estimated audio and target audio. 
They can produce good separation for non-overlapping regions but 
always suffer severe performance drops when addressing overlapping regions.
Indeed, overlap often occurs in real-world scenarios, making it one of the key issues that needs to be addressed in TSE. Wang \textit{et al.} \cite{wang2022improving} propose a TSE method utilizing timestamp information with a target-weighted loss function.
However, this system requires an additional accurate detection network, and the discriminative model still struggles in separating overlaps.

Unlike discriminative methods, generative modelling that aims to match the distribution of signals allows to approximate complex data distributions,
which have the potential to produce more natural audio.
Denoising Probabilistic Models (DPMs) have recently become increasingly popular due to their remarkable performance and reliable training. 
In particular, the intersection of DPMs and audio signal generation and synthesis tasks, such as neural vocoder \cite{chen2020wavegrad}, voice conversion \cite{popov2021diffusion}, and singing voice synthesis \cite{liu2022diffsinger}, has seen significant progress. 
For speech enhancement and separation, CDiffuSE \cite{lu2022conditional} is a DPM-based speech enhancement model designed to remove the environmental noise directly during the reverse stage of DPM, which essentially performs a discriminative task.
SGMSE \cite{10149431} is a purely generative model which demonstrates measurable advancements for speech enhancement.
Scheibler \textit{et al.} propose a source separation method \cite{scheibler2023diffusion} based on score-matching of a stochastic differential equation with higher perceptual quality than discriminative methods.
However, to the best of our knowledge, the application of DPMs in TSE has not been explored.

In this paper, we first introduce a DPM-based generative method for TSE, called DPM-TSE\footnote{Demos and source code: \href{https://jhu-lcap.github.io/DPM-TSE/}{https://jhu-lcap.github.io/DPM-TSE/}.} .
This method could better extract the target sound in the overlapping regions than discriminative methods; however, it might introduce additional noise, especially in non-target sound areas, compromising the purity of predictions.
To overcome this problem, we apply a correction method for noise schedules and sampling steps in DPM.
We conduct experiments on the FSD Kaggle 2018 dataset \cite{fonseca2017freesound} and 
objective measures show that the perceptual quality of DPM-TSE is much better than the-state-of-art discriminative models.  Subjective evaluations consistently show a preference among human listeners for the audio extracted via DPM-TSE, underscoring its heightened efficacy in extracting target sounds and eliminating irrelevant sounds.

\section{Methodology}
\label{sec:method}

\subsection{Diffusion Probabilistic Model}

Diffusion probabilistic models include a forward and a backward process. The forward process gradually adds Gaussian noise to the data, commonly based on a manually-defined variance schedule \(\beta_{1}, \ldots, \beta_{T}\). 
\begin{equation}
q\left(x_{1: T} \mid x_{0}\right):=\prod_{t=1}^{T} q\left(x_{t} \mid x_{t-1}\right) \\
\end{equation}
\vspace{-2mm}
\begin{equation}
q\left(x_{t} \mid x_{t-1}\right):=\mathcal{N}\left(x_{t} ; \sqrt{1-\beta_{t}} x_{t-1}, \beta_{t} \mathbf{I}\right)
\end{equation}

The forward process allows sampling \(x_{t}\) at an arbitrary timestep \(t\) in a closed form:
\begin{equation}
q\left(x_{t} \mid x_{0}\right):=\mathcal{N}\left(x_{t} ; \sqrt{\bar{a}_{t}} x_{0},\left(1-\bar{\alpha}_{t}\right) \mathbf{I}\right)
\end{equation}
Equivalently:
\begin{equation}
x_{t}:=\sqrt{\bar{a}_{t}} x_{0}+\sqrt{1-\bar{a}_{t}} \epsilon, \quad \text { where } \epsilon \sim \mathcal{N}(\mathbf{0}, \mathbf{I})
\end{equation}
where \(\alpha_{t}:=1-\beta_{t}\) and \(\bar{\alpha}_{t}:=\) \(\prod_{s=1}^{t} \alpha_{s}\). 

Diffusion models learn the reverse process to recover information step by step. In this way, DPM can generate new data from random Gaussian noises. When \(\beta_{t}\) is small, the reverse step is also found to be Gaussian:

\begin{equation}
p_{\theta}\left(x_{0: T}\right):=p\left(x_{T}\right) \prod_{t=1}^{T} p_{\theta}\left(x_{t-1} \mid x_{t}\right)
\end{equation}
\vspace{-2mm}
\begin{equation}
p_{\theta}\left(x_{t-1} \mid x_{t}\right):=\mathcal{N}\left(x_{t-1} ; \tilde{\mu}_{t}, \tilde{\beta}_{t} \mathbf{I}\right)
\end{equation}
where variance \(\tilde{\beta}_{t}\) can be calculated from the forward process posteriors:
$\tilde{\beta}_{t}:=\frac{1-\bar{\alpha}_{t-1}}{1-\bar{\alpha}_{t}} \beta_{t}$

In most previous DPMs, neural networks are used to predict noise \(\epsilon\), since:
\begin{equation}
\tilde{\mu}_{t}:=\frac{1}{\sqrt{\alpha_{t}}}\left(x_{t}-\frac{\beta_{t}}{\sqrt{1-\bar{\alpha}_{t}}} \epsilon\right)
\end{equation}

\subsection{Corrected Noise Schedule and Sampling Steps}

The original noise schedule commonly used in DPMs will lead to a non-zero Signal-to-noise ratio (SNR) at the last timestep \(T\), where the SNR can be calculated as:

\begin{equation}
\operatorname{SNR}(t):=\frac{\bar{\alpha}_{t}}{1-\bar{\alpha}_{t}}
\end{equation}

\begin{figure}[t] 
\centering
\includegraphics[width=8cm]{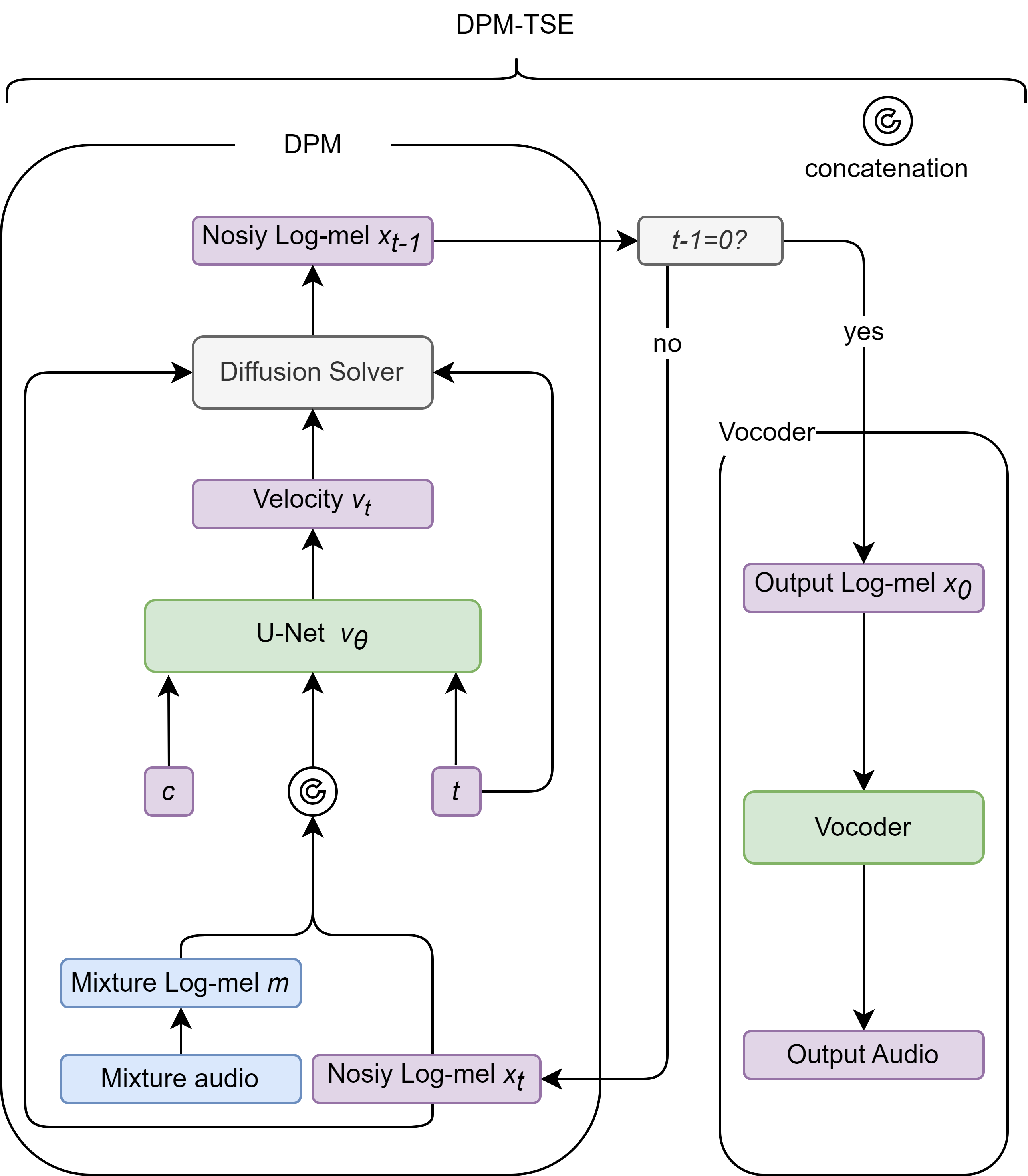}
\caption{The inference framework of Diff-TSE.}
\label{fig:f1}
\vspace{-4mm}
\end{figure}

In the field of image generation, this problem is assumed to limit the generated images to have plain medium brightness, making it difficult to generate completely dark or white image content \cite{lin2023common}. When it comes to TSE, the extracted target sound often contains many silent regions. Therefore, a non-zero terminal SNR might prevent the model from generating completely silent frames, impairing the purity and overall performance of sound extraction. Following \cite{lin2023common}, we adjust existing noise schedules to enforce zero terminal SNR by keeping \(\sqrt{\bar{\alpha}_{1}}\) unchanged, changing \(\sqrt{\bar{\alpha}_{T}}\) to zero, and linearly rescaling \(\sqrt{\bar{\alpha}_{t}}\) for intermediate \(t \in[2, \ldots, T-1]\) respectively.

When SNR is zero at the terminal step, it becomes meaningless to predict noise \(\epsilon\), as the input and output become the same. Therefore, the neural network is switched to predict velocity $v$ instead:

\begin{equation}
v_{t}:=\sqrt{\bar{\alpha}_{t}} \epsilon-\sqrt{1-\bar{\alpha}_{t}} x_{0}
\end{equation}
\begin{equation}
\epsilon=\sqrt{\bar{\alpha}_t} v+\sqrt{1-\bar{\alpha}_t} x_t
\end{equation}

According to (4) and (7), the backward process is then performed by the following functions:
\begin{equation}
x_{0}:=\sqrt{\bar{\alpha}_{t}} x_{t}-\sqrt{1-\bar{\alpha}_{t}} v_{t}
\end{equation}
\begin{equation}
\tilde{\mu}_{t}:=\frac{\sqrt{\bar{\alpha}_{t-1}} \beta_{t}}{1-\bar{\alpha}_{t}} x_{0}+\frac{\sqrt{\alpha_{t}}\left(1-\bar{\alpha}_{t-1}\right)}{1-\bar{\alpha}_{t}} x_{t}
\end{equation}

At the terminal step, the neural network with $v$ prediction now predicts the mean of the data distribution under the given conditions. Additionally, the diffusion sampler always starts from the last timestep during inference.
\vspace{-2mm}
\subsection{DPM-TSE Framework}

As shown in Figure \ref{fig:f1}, DPM-TSE comprises two modules:: a diffusion model for generating the log-mel spectrogram of the target sound conditioned on the mixture audio and the target sound token, and a neural vocoder for time-domain signal reconstruction. 
The neural network $v_{\theta}(x_{t}, m, c, t)$ with parameters ${\theta}$ in the diffusion model is used to predict velocity $v_{t}$ given the noisy target sound $x_{t}$, the audio mixture $m$, the one-hot target sound token $c$, and the corresponding diffusion step $t$,.
The diffusion step $t$ is encoded by sinusoidal position embedding \cite{vaswani2017attention}. The architecture of the diffusion network is based on U-Net \cite{ronneberger2015u} consisting of 4 downsampling blocks and 4 upsampling blocks, each of which includes 2 convolutional blocks for local feature extraction, and 2 attention blocks for capturing global time-frequency dependencies.
The HiFi-GAN vocoder \cite{kong2020hifi} trained on AudioSet \cite{gemmeke2017audio} is employed as the neural vocoder for universal audio waveform reconstruction.

\vspace{-2mm}
\section{Experimental Setups}
\label{sec:exp}
\vspace{-2mm}

\subsection{Dataset}
Following \cite{ochiai2020listen, wang2022improving}, we formulate datasets comprised of synthetic sound event mixtures using the Freesound Dataset Kaggle 2018 corpus (FSD) \cite{fonseca2017freesound}. This corpus encompasses a wide variety of 41 sound event categories ranging from human-produced sounds to musical instruments and object noises. Audio clips in the FSD have durations varying from 0.3 to 30 seconds.
We generate 10-second audio mixtures. Each mixture incorporates one target sound and 1-3 interfering sounds randomly selected from the FSD. These are then superimposed at arbitrary time points over a 10-second background noise, which we obtain from the DCASE 2019 Challenge's acoustic scene classification task \cite{mesaros2018multi}.
The signal-to-noise ratio (SNR) for each foreground sound is randomly set within a range of -5 to 10 dB. 
To optimize computational efficiency, all audio clips are down-sampled to 16 kHz. The dataset is partitioned into training, validation, and testing sets, containing 47,356, 16,000, and 16,000 samples respectively.

\vspace{-2mm}
\subsection{DPM-TSE Setups}
The default U-Net model in DPM-TSE has 4 downsampling and 4 upsampling blocks configured with 128, 256, 512, and 512 channels respectively, totaling 106.40M parameters. The larger model variant has channel configurations of 194, 384, 768, and 768, with 239.30M total parameters.
One-hot vector is applied for each target event with an embedding of 256 hidden units.
Mel-spectrogram is used as our training target of U-Net since it can provide compact acoustic features and has been successfully used in many audio tasks \cite{wang2020environmental}. 
In our experiments, we use 64-dimensional mel-spectrograms with a window size of 64 ms and a hop size of 10 ms, and we zero-pad mel-spectrograms if the number of frames is not a multiple of 4.
We use randomly segmented mel-spectrogram clips containing part of target sounds for training.
The model is trained using the Adam optimizer with a learning rate of 0.0001, a weight decay of 0.0001, batch size of 24 and 150 epochs.
The default DPM-TSE model uses corrected schedule and sampling steps and is trained with $v$ prediction. The diffusion steps and inference steps for the default DPM-TSE are 1000 and 50, and the corresponding variance is set from 0.0001 to 0.02. For the DPM-TSE with 100 diffusion steps and 30 inference steps, the variance is set from 0.0001 to 0.06.
\vspace{-6mm}
\subsection{Baselines}

We utilize two latest TSE models, namely WaveFormer and Tim-TSENet with the same settings of their original implementations, as our baselines. WaveFormer \cite{veluri2023real} is a time-domain source separation model based on Conv-TasNet \cite{luo2019conv}, which incorporates transformer blocks. Tim-TSENet \cite{wang2022improving} proposes an STFT-based TSE model with the similar masking strategy used in Conv-TasNet. 
For fair comparision, we also tried mel-spectrogram-based Tim-TSNet and STFT-based DPM-TSE \footnote{Results of other attempts: \href{https://jhu-lcap.github.io/DPM-TSE/}{https://jhu-lcap.github.io/DPM-TSE/}.}.
The Mel-spectrogram-based Tim-TSENet exhibited a performance degradation due to the difficulty of introducing a time-domain loss function using inverse STFT as in the original Tim-TSENet. Meanwhile, the STFT-based DPM-TSE suffered from significant performance degradation and excessive computational complexity.

\begin{table*}[t]
  \caption{Objective and subjective scores with their 95\% confidence intervals. ViSQOL-T and CDPAM-T are calculated with the target sound regions, while other scores are calculated with the whole audio.}
  \label{tab:result1}
  \footnotesize
  \centering
  \begin{tabular}{lcc|cc|cc}
    \hline
    \textbf{Method}&\textbf{ViSQOL $\uparrow$}&\textbf{CDPAM $\downarrow$}&\textbf{ViSQOL-T $\uparrow$}&\textbf{CDPAM-T $\downarrow$}&\textbf{Extraction $\uparrow$} &\textbf{Purity $\uparrow$}\\
    \hline
    WaveFormer \cite{veluri2023real} & $1.96\pm0.05$ & $0.38\pm0.02$ & $1.78\pm0.05$ & $0.50\pm0.02$ & $3.38\pm0.17$ & $2.61\pm0.19$\\
    TSENet \cite{wang2022improving} & $2.32\pm0.05$ & $0.31\pm0.02$ & $2.04\pm0.05$ & $0.42\pm0.02$ & $3.80\pm0.18$ & $3.19\pm0.21$ \\
    \textbf{DPM-TSE}  & \boldsymbol{$2.53\pm0.05$} & \boldsymbol{$0.22\pm0.01$} & \boldsymbol{$2.18\pm0.05$} & \boldsymbol{$0.38\pm0.03$} & \boldsymbol{$4.19\pm0.14$} &  \boldsymbol{$3.74\pm0.18$}\\
    \hline
  \end{tabular}
\end{table*}

\begin{figure*}[t] 
\centering
\includegraphics[width=16cm]{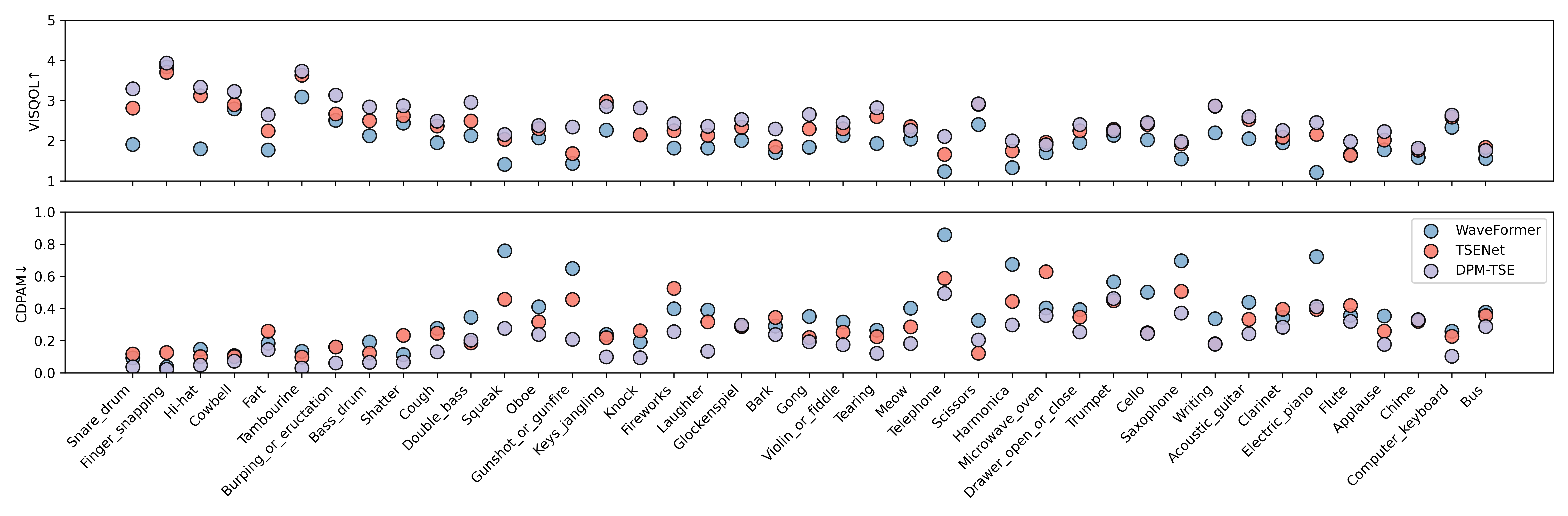}
\caption{Distribution of objective performance by sound category in ascending order of average sound event duration.}
\label{fig:f2}
\vspace{-2mm}
\end{figure*}

\begin{table}[t]
  \caption{Results of ablation study on DPM-TSE based on objective scores with their 95\% confidence intervals.}
  \label{tab:result2}
  \footnotesize
  \centering
  \begin{tabular}{ccc|cc}
    \hline
    \textbf{Model}&\textbf{Schedule}&\textbf{Steps}&\textbf{ViSQOL $\uparrow$}&\textbf{CDPAM $\downarrow$}\\
    \hline
    Base & Default & 1000/50 & $2.39\pm0.06$ & $0.34\pm0.02$\\
    Base & Corrected & 100/30 & $2.43\pm0.05$ & $0.25\pm0.01$ \\
    \textbf{Base} & \textbf{Corrected} & \textbf{1000/50} & \boldsymbol{$2.53\pm0.05$} & \boldsymbol{$0.22\pm0.01$} \\
    Large & Corrected & 1000/50  & $2.38\pm0.05$ & $0.24\pm0.01$\\
    \hline
  \end{tabular}
  \vspace{-3mm}
\end{table}

\vspace{-2mm}
\subsection{Evaluation Metrics}
The primary objective of this study is to enhance the auditory quality of the output generated through TSE. As such, we opt for a  several perceptual evaluations and subjective assessment to gauge the performance of the target sound extraction models. We steer away from relying solely on objective measures such as Signal-to-Distortion Ratio (SDR) \cite{vincent2006performance}, since existing objective metrics for source separation are imperfect proxies for human auditory perception, as highlighted in previous research \cite{le2019sdr, cartwright2018crowdsourced}.

\noindent \textbf{Objective metrics:}
We use two automatic evaluation functions:
\noindent (1) \textbf{ViSQOL} \cite{chinen2020visqol} is an algorithm originally designed to predict the quality of speech signals, and has since been adapted to assess the quality of audio signals by approximating human perceptual responses based on five-scaled mean opinion scores.
\noindent (2) \textbf{CDPAM} \cite{manocha2021cdpam} is a perceptual audio metric based on deep neural network that correlates well with human subjective ratings across sound quality assessment tasks, measuring audio similarity by distance of deep features.

\noindent \textbf{Human evaluation: }
For subjective evaluation, 15 participants with recording or music production experiences were recruited to evaluate the listening perceptual quality of audios predicted by different TSE models.
We randomly selected 1 sample from 41 sound categories from the test set. 
Each subject was asked to evaluate 20 randomly assigned audio pairs for each model, and each audio pair contains both a ground truth and a model prediction for the extracted sound.
They were given two questions for each audio pair:
(1) \textbf{Extraction: Does the generated audio contain everything from the reference audio?}
Rating from 1 to 5, where 1 means that the content of the reference audio cannot be heard at all in the generated audio, and 5 means that the second segment completely contains everything from the reference audio
(2) \textbf{Purity: Does the generated audio only have the sound from the reference audio?}
Rating from 1 to 5, where 1 means that it is pretty obvious that the generated audio has a lot of sound that the reference audio doesn't have, and 5 means that the generated audio only has the sound corresponding to the reference audio and other sounds cannot be detected.
 \vspace{-3mm}
\section{Results}

The results in Table \ref{tab:result1} demonstrate that DPM-TSE achieves the best performance in both subjective and objective experiments. The key observations include: (1) DPM-TSE has a promising performance in localizing and recovering target sound. (2) DPM-TSE shows a significant advantage of producing cleaner target sound, while Tim-TSENet and WaveFormer fail to remove non-target sound very well, especially in regions where there is overlapping between target and other sounds. 
In Fig. \ref{fig:f2}, we further explore the performance of target sound extraction in different sounds categories based on objective metrics. The three models simultaneously show good results in short-duration events (like finger snapping, tambourine, cowbell and hi-hat) while performs poor for long-duration complex events (like bus, saxophone, chime and flute).
CDPAM and ViSQOL has similar distribution across the majority of classes.
In most categories, DPM-TSE demonstrates pronounced advantages. 

In addition, we conduct ablation study on noise schedule methods, number of training and inference steps, and model size.
As shown in Table \ref{tab:result2},
the proposed corrected noise schedule significantly improves the performance of the model. We find that the DPM-TSE using the original noise schedule produces additional noise, which is prominently noticeable in non-target sound regions.
The DPM-TSE with larger model shows a performance degradation, which may be limited by the size of dataset.
Comparing 100 training steps with 1000 training step,
we find that the DPM-TSE model with fewer diffusion and inference steps still achieves relatively good performance and can be used in situations where faster inference is preferred.

 \vspace{-2mm}
\section{Conclusion}
\label{sec:conclusions}
In this paper, we propose a DPM-based generative method for TSE, which is quite effective at extracting target sounds and removing irrelevant sounds.
In future works, our focus will pivot towards (1) enhancing the sampling speed of DPM-TSE and (2) delving into innovative avenues including zero-shot TSE and text-guided TSE and audio editing techniques.

\bibliographystyle{IEEEbib}
\footnotesize
\bibliography{refs}

\end{document}